\documentclass[showpacs,twocolumn,APS]{revtex4}

\usepackage{graphicx}
\usepackage{dcolumn}
\usepackage{amsmath}
\makeatletter
\def\btt#1{\texttt{\@backslashchar#1}}
\DeclareRobustCommand\bblash{\btt{\@backslashchar}} \makeatother

\begin{document}
\title[]{Uniform density static fluid sphere in Einstein-Gauss-Bonnet gravity and its universality}

\author{Naresh Dadhich}
\affiliation{Inter-University Centre for Astronomy \& Astrophysics,\\ Post Bag 4 Pune 411 007, India}
\author{Alfred Molina}
\affiliation{Departament de F\'{\i}sica Fonamental, Universitat de Barcelona, Spain}
\author{Avas Khugaev}
\affiliation{Institute of Nuclear Physics, Tashkent, 100214, Uzbekistan}
\date{\today}

\begin{abstract}

In Newtonian theory, gravity inside a constant density static sphere is independent of spacetime dimension. Interestingly this general result is also carried over to Einsteinian  as well as higher order Einstein-Gauss-Bonnet (Lovelock) gravity notwithstanding their nonlinearity. We prove that the necessary and sufficient condition for universality of the  Schwarzschild interior solution describing a uniform density sphere for all $n\geq4$ is that its density is constant. 
\end{abstract}

 
\maketitle
\section{Introduction}

In Newtonian gravity, the gravitational potential at any point inside a fluid sphere is given by $-M(r)/r^{n-3}$ for $n\geq4$ dimensional spacetime. Now $M(r)= \int \rho r^{n-2}dr$ which for constant density will go as $\rho r^{n-1}$ and then the potential will go as $\rho r^{n-1}/r^{n-3} = \rho r^2$ and is therefore independent of the dimension. This is an interesting general result: for the uniform density sphere, gravity has the universal character that it is independent of the dimension of spacetime. It is then a natural question to ask, Does this result carry over to Einsteinian gravity? In general relativistic language it is equivalent to ask, Does  Schwarzschild interior solution that describes the uniform density sphere in four dimensions remain good for all $n\geq4$? The main purpose of this paper is to show that it is indeed the case not only for Einstein gravity but also for higher order Einstein-Gauss-Bonnet (Lovelock) gravity. It is remarkable that this general feature holds true notwithstanding the highly nonlinear character of the theory. 

In static spherically symmetric fluid spacetime, we have two equations to handle: one is for density which easily integrates to give $g_{rr}$,and the other is the pressure isotropy equation determining $g_{tt}$. So long as density remains constant,the former equation will always integrate to give $g_{rr}$ in all dimensions with constant density redefined. Then we just need to make the latter equation free of dimension $n$ so that the constant density Schwarzschild interior solution becomes universally true for all $n$. In particular it turns out that the universality condition indeed implies constant density. Thus constant density is a  necessary and sufficient condition for universality of the Schwarzschild interior solution for $n\geq4$ not only for Einstein but also for Einstein-Gauss-Bonnet (EGB) theory. 

Higher dimension is a natural playground for string theory and string inspired investigations (see a comprehensive review \cite{emparan}). The most popular studies have been of higher dimensional black holes \cite{tang-mayper-chong} with a view to gain greater and deeper insight into quantum phenomena, black hole entropy and the well-known AdS/CFT correspondence \cite{stro-mal-witt}. There have also been studies of fluid spheres in higher dimensions \cite{krori-shen-ponce}. We shall, however, focus on the universal character of constant density solution in Einstein and EGB theory and its matching with the corresponding exterior solution. The paper is organized as follows. In the next section, we establish the universality of the uniform density solution for Einstein and EGB theories and demonstrate the matching with an exterior solution for the five-dimensional Gauss-Bonnet black hole. We conclude with a discussion. 

\section{Uniform density sphere}     
\subsection{Einstein case}
We begin with the general static spherically symmetric metric given by 
\begin{equation}
ds^2= e^\nu dt^2 - e^\lambda dr^2 - r^2d\Omega_{n-2}^2
\end{equation}
where $d\Omega_{n-2}^2$ is the metric on a unit $(n-2)$-sphere. For the Einstein equation in the natural units ($8\pi G=c=1$),
\begin{equation}
G_{AB} = R_{AB} - \frac{1}{2} R g_{AB} = - T_{AB}
\end{equation}
and for perfect fluid, $T_A^B = diag(\rho, -p, -p, ..., -p)$,
we write 
\begin{equation}
e^{-\lambda}(\frac{\lambda^{\prime}}{r} - \frac{n-3}{r^2}) + \frac{n-3}{r^2} = \frac{2}{n-2}\rho  \label{density}
\end{equation}
\begin{equation}
e^{-\lambda}(\frac{\nu^{\prime}}{r} + \frac{n-3}{r^2}) - \frac{n-3}{r^2} = \frac{2}{n-2}p 
\end{equation}
and the pressure isotropy is given by 
\begin{eqnarray}
e^{-\lambda}(2\nu^{\prime\prime} + \nu^{\prime^2} - \lambda^{\prime}\nu^{\prime} - 2\frac{\nu^{\prime}}{r}) \nonumber \\
- 2(n-3)(\frac{e^{-\lambda}\lambda^{\prime}}{r} 
+ 2\frac{e^{-\lambda}}{r^2} - \frac{2}{r^2}) = 0  \label{isotropy}. 
\end{eqnarray}
Let us rewrite this equation in a form that readily yields the universal character of the Schwarzschild interior solution for all $n\geq4$,  
\begin{eqnarray} 
e^{-\lambda}(2\nu^{\prime\prime} + \nu^{\prime^2} - \lambda^{\prime}\nu^{\prime} - 2\frac{\nu^{\prime} + \lambda^{\prime}}{r} - \frac{4}{r^2}) + \frac{4}{r^2} \nonumber \\ 
- 2(n-4) \Bigl((n-1)(\frac{e^{-\lambda}}{r^2} - \frac{1}{r^2}) + \frac{2\rho}{n-2} \Bigr) = 0. \label{iso}
\end{eqnarray}
We now set the coefficient of $(n-4)$ to zero so that the equation remains the same for all $n\geq4$. This then straightway determines $e^{-\lambda}$ without integration and it is given by 
\begin{equation}
e^{-\lambda} = 1 - \rho_0 r^2 \label{sol} 
\end{equation} 
where $\rho_0=2\rho/{(n-1)(n-2)}$. This when put in Eq. (\ref{density}) implies constant density. We thus obtain $\rho=const.$ as the neceessary condition for universality of the isotropy equation for all $n\geq4$. The sufficiency of constant density is obvious from the integration of Eq. (\ref{density}) for $\rho=const$, giving the same solution as above where a constant of integration is set to zero for regularity at the center. Thus constant density is a necessary and sufficient condition for universality of field inside a fluid sphere, i.e. independent of spacetime dimension. An alternative identification of constant density is that the gravitational field inside a fluid sphere is independent of spacetime dimension $\geq4$. This universal property is therefore true if and only if density is constant. 

As is well known, Eq. (\ref{iso}) on substituting Eq. (\ref{sol}) admits the general solution as given by  
\begin{equation}
e^{\nu/2} = A + Be^{-\lambda/2} \label{sol2}
\end{equation}
where $A$ and $B$ are constants of integration to be determined by matching to the exterior solution. This is the Schwarzschild interior solution for a constant density sphere that is independent of the dimension except for a  redefinition of the constant density as $\rho_0$. This proves the universality of the Schwarzschild interior solution for all $n\geq4$. 

The Newtonian result that gravity inside a uniform density sphere is independent of spacetime dimension is thus carried over to general relativity as well despite nonlinearity of the equations. That is, Schwarzschild interior solution is valid for all $n\geq4$. Since there exist more general actions like Lovelock polynomial and $f(R)$ than the linear Einstein-Hilbert, it would be interesting to see whether this result would carry through there as well. That is what we take up next.  

\subsection{Gauss-Bonnet(Lovelock) case} 
There is a natural generalization of Einstein action to  Lovelock action that is a homogeneous polynomial in Riemann curvature with Einstein being the linear order. It has the remarkable property that on variation it still gives the second order quasilinear equation that is its distinguishing feature. The higher order terms make a nonzero contribution in the equation only for dimension $\geq5$. The quadratic term in the polynomial is known as Gauss-Bonnet, and for that we write the action as 
\begin{equation} 
\label{action}
S=\int d^nx\sqrt{-g}\biggl[\frac{1}{2}(R-2\Lambda+\alpha{L}_{GB}) \biggr]+S_{\rm matter},
\end{equation}
where $\alpha$ is the GB coupling constant and all other symbols have their usual meaning. 
The GB Lagrangian is the specific combination of Ricci scalar, Ricci, and Riemann curvatures, and it is given by 
\begin{equation}
{L}_{GB}=R^2-4R_{AB}R^{AB}+R_{ABCD}R^{ABCD}.
\end{equation}
This form of action is known also to follow from the low-energy limit of heterotic superstring theory~\cite{gross}. 
In that case, $\alpha$ is identified with the inverse string tension and is positive definite, which is also required for the stability of Minkowski spacetime.

The gravitational equation following from the action (\ref{action}) is given by 
\begin{equation}
G^A_B +\alpha H^A_B = - T^A_B, \label{beq}
\end{equation}
where 
\begin{eqnarray}
&& H_{AB}\equiv 2\Bigl[RR_{AB}-2R_{AC}R^C_B-2R^{CD}R_{ACBD}
\nonumber\\
&& \hspace*{4em}
 +R_{A}^{~CDE}R_{BCDE}\Bigr]
-{\frac12}g_{AB}{L}_{GB}.\label{def-H}
\end{eqnarray}
Now density and pressure would read as follows:
\begin{eqnarray}
\rho = \frac{(n-2)e^{-\lambda}}{2r^2} \Bigl(r\lambda^\prime - 
(n-3)(1-e^{\lambda})\Bigr)+\qquad\qquad\nonumber \\ 
+\frac{(n-2) e^{-2\lambda}\tilde\alpha}{2r^4}(1-e^{\lambda}) 
\Bigl(-2r\lambda^\prime + (n-5)(1-e^{\lambda})\Bigr) 
\label{rho-gb} 
\end{eqnarray}
\begin{eqnarray}
p = \frac{(n-2)e^{-\lambda}}{2r^2} \Bigl(r\nu^\prime + (n-3)(1-e^{\lambda})
\Bigr)-\qquad\qquad \nonumber \\
- \frac{(n-2) e^{-2\lambda}\tilde\alpha}{2r^4}(1-e^{\lambda}) \Bigl(2r\nu^\prime 
+ (n-5)(1-e^{\lambda})\Bigr).
\label{p-gb}
\end{eqnarray}

The analogue of the isotropy Eq. (\ref{iso}) takes the form 
\begin{equation}
I_{GB} \equiv \biggl(1+\frac{2\tilde\alpha f}{r^2}\biggr){I_E}+
\frac{2\tilde\alpha}{r}\biggl(\frac{f}{r^2}\biggr)^{\prime}
\biggl[r\psi^\prime + \frac{f}{1-f}\psi\biggr] = 0 
\label{iso-gb}
\end{equation}
where $\psi=e^{\nu/2}, e^{-\lambda}=1-f, \tilde\alpha=(n-3)(n-4)\alpha$ and $I_E$ is given by the left-hand side (LHS)  of Eq. (\ref{isotropy}), 
\begin{eqnarray}
I_E \equiv \frac{(1-f)}{\psi}\biggl\{\psi^{\prime\prime} -
\biggl(\frac{f^{\prime}}{2(1-f)}+\frac{1}{r}\biggr)\psi^{\prime} - 
\qquad\qquad\nonumber  \\
- \frac{(n-3)}{2r^2(1-f)}(rf^{\prime} - 2f)\psi\biggr\}. \label{isop}
\end{eqnarray}

From Eq. (\ref{rho-gb}), we write 
\begin{equation}
(\tilde\alpha r^{n-5}f^2 + r^{n-3}f)^{\prime} = \frac
{2}{n-2}\rho r^{n-2}
\end{equation}
which integrates for $\rho=const.$ to give  
\begin{equation}
\tilde\alpha r^{n-5}f^2 + r^{n-3}f = \rho_0r^{n-1} + k	
\end{equation}                                    
where $k$ is a constant of integration that should be set to zero for regularity at the center and $2\rho/(n-1)(n-2)=\rho_0$ as defined earlier. Solving for $f$, we get 
\begin{equation}
e^{-\lambda} = 1 - f = 1 - \rho_{0GB}r^2 \label{sol-gb}	
\end{equation}
where 
\begin{equation} 
\rho_{0GB} = \frac{\sqrt{1 + 4\tilde\alpha\rho_0} - 1}{2\tilde\alpha}.
\end{equation}
So the solution is the same as in the Einstein case and the  appropriate choice of sign is made so as to admit the limit $\alpha\rightarrow 0$ yielding the Einstein $\rho_0$ (the other choice would imply $\rho_{0GB}<0$ for positive $\alpha$). This, when substituted in the pressure isotropy Eq. (\ref{iso-gb}), would lead to $I_E=0$ in Eq. (\ref{isop}) yielding the solution (\ref{sol2}) as before. This establishes sufficient condition for universality. 

For the necessary condition, we have from Eqn (\ref{iso-gb}) that either 
\begin{equation}
\Bigl(\frac{f}{r^2}\Bigr)^\prime = 0
\end{equation} 
or 
\begin{equation}
r\psi^\prime + \frac{f}{1-f}\psi = 0
\end{equation} 

The former straightway leads with the use of Eq. (17) to the same constant density solution (\ref{sol-gb}) and $I_E=0$ integrates to Eq. (\ref{sol2}) as before. This shows that universality implies constant density as the necessary condition. For the latter case, when Eq. (22) is substituted in Eq. (\ref{isop}) and $I_E = 0$ is now solved for $\lambda$, we again obtain the same solution (\ref{sol-gb}). Equation (17) again implies $\rho=const$ as the necessary condition. Now $\psi$ is determined by Eq. (22), which means the constant $A$ in solution (\ref{sol2}) must vanish. Then the  solution turns into de Sitter spacetime with $\rho=-p=const.$ which is a particular case of Schwarzschild solution. This is, however, not a bounded finite distribution. 

Thus universality and finiteness of a fluid sphere uniquely characterize the Schwarzschild interior solution for Einstein as well as for Einstein-Gauss-Bonnet gravity. That is, gravity inside a fluid sphere of finite radius is universal, i.e it is true for all $n\geq4$ if and only if the density is constant and it is described by the Schwarzschild interior solution. It is only the constant density that gets redefined in terms of $\rho_0$ and $\rho_{0GB}$. If we relax the condition of finiteness, it is de Sitter spacetime with $\rho=-p=const$. 

Our entire analysis is based on the two equations (\ref{iso-gb}) and (17). Let us look at GB contributions in them. In the former, there is a multiplying factor to the Einstein second order differential operator $I_E$ and another term with the factor $\tilde\alpha (f/r^2)^\prime$. This indicates that the contributions of higher orders in Lovelock polynomial will obey this pattern to respect quasilinearity of the equation. The higher orders will simply mean inclusion of the corresponding couplings in the multiplying factor as well as in the second term appropriately while the crucial entities, $I_E$ and $(f/r^2)^\prime$ on which the proof of the universality of Schwarzschild solution hinges remain intact. On the other hand, Eq. (17) is quadratic in $f$ for the quadratic GB action, which means the degree of $f$ is tied to the order of the Lovelock polynomial. It essentially indicates that as $\rho_{0GB}$ is obtained from a quadratic algebraic relation, similarly in higher order its analogue will be determined by the higher degree algebraic relation. The solution will always be given by Eq. (\ref{sol}). Thus what we have shown  explicitly for EGB will go through for the general Einstein-Lovelock gravity. 

Since Eqs. (15-17) owe their form and character to quasilinearity of the EGB equation, hence the carrying through of the Newtonian result of universality of gravity inside a uniform density fluid sphere critically hinges on quasilinearity.  Thus this general result will not go through in theories like $f(R)$ gravity which do not in general respect quasilinearity. It could in a sense be thought of as yet another identifying feature of Einstein-Lovelock gravity. 

Let us now also indicate an itriguing and unusual feature of GB(Lovelock) gravity. What happens if the multiplying  factor, $1+2\tilde\alpha f/r^2 = 0$ in Eq. (\ref{iso-gb})? Then the entire equation becomes vacuous, leaving $\psi$ completely free and undetermined while $e^{-\lambda} = 1 + r^2/2\tilde\alpha$. This leads to $p = -\rho = (n-1)(n-2)/8\tilde\alpha$, which is an anti-de Sitter distribution for $\alpha\ge0$. This is a special prescription where density is given by GB coupling $\alpha$. There is no way to determine $\psi$, and so we have a case of genuine indeterminacy of the metric. It is because GB(Lovelock) contributes such a multiplying factor involving ($\alpha,~ r,~ f$) to sceond order quasilinear operator, which could be set to zero and thereby annul the equation altogether. Such a situation has been studied in the Kaluza-Klein split-up of six-dimensional spacetime into the usual $M^4$ and $2$-space of constant curvature in EGB theory ~\cite{mada}. It gave rise to a black hole from pure curvature where the equations split-up into a four-dimensional part and a scalar constraint from an extra- dimensional part. As here by fine-tuning $\alpha, \Lambda$, and the constant curvature of the $2$-space, the four-dimensional part was turned vacuous, and then the metric was, however, determined by the remaning single scalar equation. This was because for vacuum (the null energy condition implies $\nu+\lambda =0$ in our notation), there was only one free parameter to be determined for which there was still a scalar constraint equation. The solution of that gave the black hole without matter support on $M^4$ ~\cite{mada}. In contrast, here we have two metric functions to be determined and there is only one equation remaining after the fine-tuning of density with $\alpha$. Thus one metric function will have to remain undetermined. As argued above, the form of Eq. (\ref{iso-gb}) will be generic for the Lovelock system, and hence this kind of indeterminacy under the fine-tuning of parameters will also be generic. 

\subsection{Matching with the exterior} 
Now we would like to demonstrate matching of the interior with the corresponding exterior five-dimensional Gauss-Bonnet  black hole solution \cite{boul}. In the interior,  pressure is given by 
\begin{eqnarray}
&&\hspace*{-3em} p=\frac{3}{4\alpha}(1-\mu )
\biggl[ 1- \frac{\mu}{1+\frac{2A\sqrt\alpha}{B\sqrt{r^2(1-\mu )+4\alpha}}} 
\biggr] 
\end{eqnarray}
where 
\begin{equation}
\mu = \sqrt{1 + 8\alpha \rho_{0GB}}.
\end{equation}
At the boundary, $r=r_\Sigma$, pressure vanishes, which is equivalent to the continuity of $g_{tt}^\prime$, and that is what we shall employ. Besides this, the metric should be continuous across $r_\Sigma$. 
The five-dimensional Gauss-Bonnet black hole is given by the metric \cite{boul}, 
$$ds^2= F(r)dt^2 - \frac{dr^2}{F(r)} - r^2(d\theta^2+\sin^2(\theta)( d\varphi^2+\sin^2(\psi) d\psi^2))$$
where 
$$F(r) = 1 + \frac{r^2}{4\alpha}(1 -\sqrt{1 + 8M\alpha/r^4}).
$$
Now matching $g_{rr}$ means $[g_{rr}]_\Sigma=0$ which after appropriate substitutions determines the mass enclosed inside the radius $r_\Sigma$,
\begin{equation}
M= \frac16\rho_{0GB} r_\Sigma^4. 
\end{equation} 
Further $[g_{tt}]_\Sigma=0$ and $[g_{tt}^\prime]_\Sigma=0$ determine the constants, 
\begin{equation}
A=(1-B)\sqrt{1-\rho_{0GB}r_\Sigma^2}
\end{equation} 
and
\begin{equation}
B=-(1 + \frac{8\alpha M}{r_\Sigma^4})^{-1/2}.
\end{equation} 
This completes the matching of the interior and exterior solutions. 

\section{Discussion}

We have established that the gravitational field inside a constant density fluid sphere has a universal character for spacetime dimensions $\geq4$. This is true not only for Einstein-Hilbert action but also for the more general  Lovelock action which is a homogeneous polynomial in Riemann curvature. We  have explicitly shown this for the linear Einstein and the quadratic Gauss-Bonnet cases and have argued that the proof would go through for the general Lovelock polynomial. That is, the Schwarzschild interior solution describing the gravitational field of the constant density sphere is true for all spacetime dimensions $\geq4$ for Einstein as well as for higher order Einstein-Lovelock polynomial gravity. It turns out that the  necessary and sufficient condition for the universality of fluid sphere is that its density must be constant. Equivalently, universality uniquely characterizes the Schwazschild interior solution for a fluid sphere of finite radius.   

This result is obvious but perhaps not much noticed in Newtonian gravity as argued in the opening of the paper. It is, however, not so for Einstein-Lovelock gravity because of its highly nonlinear character. Yet it is carried through because the equation of motion still remains second order quasilinear. It is this feature that carries the general character of the solution into higher order gravity. Clearly it would not in general be carried along for non-quasilinear theory like $f(R)$ gravity. Apart from Lovelock's original derivation of the action \cite{lov}, there are two other characterizations of Lovelock action \cite{dad,irani}. In \cite{dad}, the identifying feature is the existence of the  homogeneous polynomial  in curvatures analogus to the Riemann curvature whose trace of the Bianchi derivative yields the corresponding analogue of the Einstein tensor in the equation while for \cite{irani} it is the requirement that both metric and Palitini variations give the same equation of motion. Here we have yet another identifying property of Einstein-Lovelock gravity. Also it exhibits that the obvious Newtonian result is carried through in higher order nonlinear theories. Universality characterizes uniform density for the static fluid sphere.  
 
The main aim of such investigations is essentially to probe and identify universal features of gravity for greater understanding and insight. Such universal features also  provide discerning criteria for competing genralizations of Einstein gravity.  
   
\acknowledgments
This work was initiated by ND and AM during their visit to the  University of Kwa-Zulu Natal, Durban, and they would like to warmly thank Professor Sunil Maharaj for the wonderful hospitality. AK gratefully acknowledges TWAS for support while visiting IUCAA and also thanks IUCAA for warm hospitality which facilitated this collaboration. AM also acknowledges financial suport from the Ministerio de Educaci\'on, Grant No. FIS2007/63 and from the Generalitat de Catalunya, 2009SGR0417 (DURSI). Finally we also thank the  anonymous referee for the constructive criticism that has substantially improved clarity and precision.

\end{document}